\documentclass[preprint,prb,amssymb,showpacs]{revtex4}
\usepackage{graphicx}

\begin{document}

\title{Antilocalization in gated 2D
quantum well structures with composition gradient }

\author{A.~V.~Germanenko}
\author{G.~M.~Minkov}
\author{O.~E.~Rut}
\author{A.~A.~Sherstobitov}

\email{Alexander.Germanenko@usu.ru}

\affiliation{Institute of Physics and Applied Mathematics, Ural
State University, 620083 Ekaterinburg, Russia}

\author{B.~N.~Zvonkov}

\affiliation{Physical-Technical Research Institute, University of
Nizhni Novgorod, \\  Nizhni Novgorod 603600, Russia }


\begin{abstract}
Low-field magnetoconductivity caused by the quantum interference
is studied in the gated 2D quantum well structures with the
composition gradient. It is shown that the Dresselhaus mechanism
well describes an antilocalization minimum on the
conductivity-magnetic field curve.
\end{abstract}

\maketitle

\section{Introduction}    

An asymmetry of the quantum well leads to splitting of the energy
spectrum due to spin-orbit interaction. The inversion asymmetry of
the crystal field results in linear and cubic in quasimomentum
terms in the energy spectrum and thus leads to so called
Dresselhaus spin-orbit splitting.\cite{Mindress} The dispersion
law for the heterostructure grown in [001] direction looks in this
case as follows
\begin{equation}
E(k,\varphi)={k^2 \over 2m} \pm
\sqrt{\Omega_1^2+\Omega_3^2+2\Omega_1\Omega_3 \cos(4\varphi)},
 \label{eq1}
\end{equation}
where  $\Omega_1=\gamma k \left(\langle k_z^2 \rangle -
k^2/4\right)$, $\Omega_3=\gamma k^3/4$, $k^2=k_x^2+k_y^2$,
$\tan\varphi=k_x/ k_y$, $\langle k_z^2 \rangle$ is the mean square
of electron momentum in the direction  perpendicular to 2D plane,
and  $\gamma$ is the constant of spin-orbit interaction, which
depends on the band parameters of the bulk material.

An asymmetry of smooth electrostatic potential in the
perpendicular to the 2D plane direction due to Schottky barrier or
due to asymmetry in disposition of doping layers gives linear term
or, so called, the Rashba term:\cite{Bychkov}
\begin{equation}
E(k)={k^2 \over 2m} \pm \Omega_R,
 \label{eq2}
\end{equation}
where $\Omega_R=\alpha |k|$. The value of parameter $\alpha$ can
be found within framework of $kP$-formalism:
\begin{equation}
\alpha= {P^2 \over 3}\left< \Psi(z)\left|{d\over dz}\left( {1
\over E_F-E_{\Gamma_7}(z)}-{1 \over
E_F-E_{\Gamma_8}(z)}\right)\right|\Psi(z)\right>,\label{eq3}
\end{equation}
where $\Psi(z)$ is the wave function for the confined electrons,
$P$ is the momentum matrix element, $E_F$ is the Fermi energy, and
$E_{\Gamma_7}(z)$ and $E_{\Gamma_8}(z)$ are the positions of the
band edges energies for $\Gamma_7$ and $\Gamma_8$ valence bands,
respectively, at position $z$.

\begin{figure}
\includegraphics[width=10cm,clip=true]{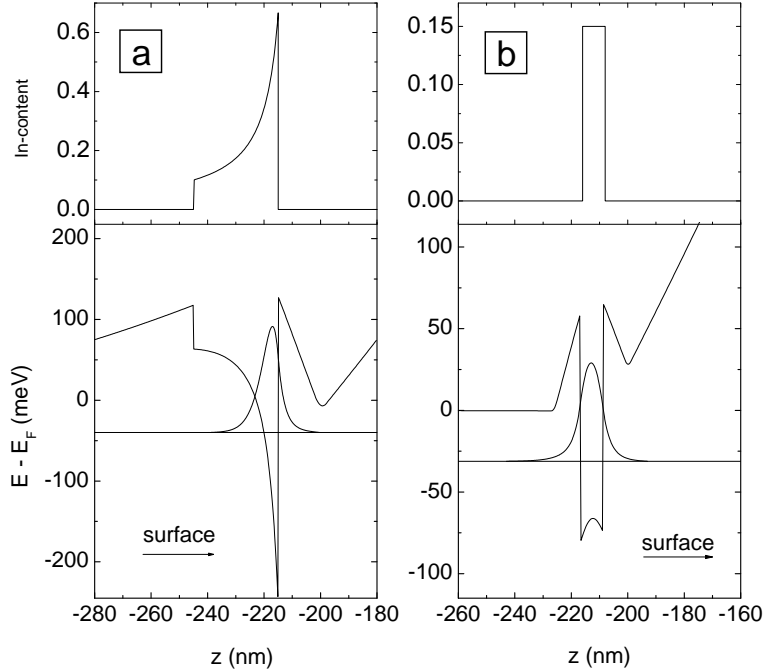}
\caption{Indium distribution (upper panels) and energy diagrams
(lower panels) for structures 3635 (a) and 3512 (b).} \label{f1}
\end{figure}

As seen from Eq.~(\ref{eq3}) an appropriate disposition of the
doping layers and quantum well, in principle, gives a possibility
to engineer the value of the spin splitting. Another way to change
the splitting is a variation of composition within the quantum
well in the growth direction.

One of suitable way to study the spin splitting of the energy
spectrum at zero-magnetic field is analysis of the weak
antilocalization. This paper is devoted to experimental study of
the low magnetic field positive magnetoresistance caused by the
spin relaxation in GaAs/In$_{x}$Ga$_{1-x}$As/GaAs quantum wells.
\section{Experimental details}

The quantum well GaAs/In$_{x}$Ga$_{1-x}$As/GaAs heterostructures
were grown by metal-organic vapor phase epitaxy on semi-insulator
GaAs substrate. Two types of heterostructures were measured.

Asymmetric heterostructure 3635 consists of a 0.3~$\mu$m-thick
undoped GaAs buffer layer, a 30~nm In$_{x}$Ga$_{1-x}$As well, a
15~nm spacer of undoped GaAs, a Si $\delta$-layer   and  200~nm
cap layer of undoped GaAs. The concentration of In within the
quantum well varies from $0.1$ to $0.6$ from the buffer to cap as
$0.6/[6-0.17(z+245)]$, where $z$ is the coordinate perpendicular
to the quantum well plane, measured in nanometers [upper panel in
Fig.~\ref{f1}(a)]. The energy diagram calculated for this
structure is presented in lower panel in Fig.~\ref{f1}(a) The
electron density $n$ and mobility $\mu$ are the following:
$n=8\times 10^{15}$ m$^{-2}$, $\mu=2.4$ m$^2$/Vs.

Heterostructures 3512 and Z76 are symmetric. The heterostructure
3512 consists of 0.5 mkm-thick undoped GaAs epilayer, a Sn
$\delta$-layer, a 9 nm spacer of undoped GaAs, a 8 nm
In$_{0.2}$Ga$_{0.8}$As well, a 9 nm spacer of undoped GaAs, a Sn
$\delta$-layer, and a 300 nm cap layer of undoped GaAs. The
electron density and mobility are the following: $n=9.5\times
10^{15}$ m$^{-2}$, $\mu=1.4$ m$^2$/Vs. The energy diagrams for
structure 3512 is presented in Fig.~\ref{f1}(b). The structure Z76
is analogous. The only difference is thickness of spacer between
the Sn $\delta$-layers and quantum well, which is 12~nm. The
parameters of this structure are $n=6.2\times 10^{15}$ m$^{-2}$
and $\mu=2.4$ m$^2$/Vs.

The samples were mesa etched into standard Hall bars and then an
Ag or Al gate electrode was deposited by thermal evaporation onto
the cap layer through a mask. Varying the gate voltage $V_g$ we
were able to control the density and conductivity of electron gas
in the quantum well. The low field magnetoresistance was measured
with step $10^{-2}$~mT within the temperature range $0.45-5.0$~K.

\section{Results and discussion}
\label{sec:res}

\begin{figure}[t]
\includegraphics[width=7cm,clip=true]{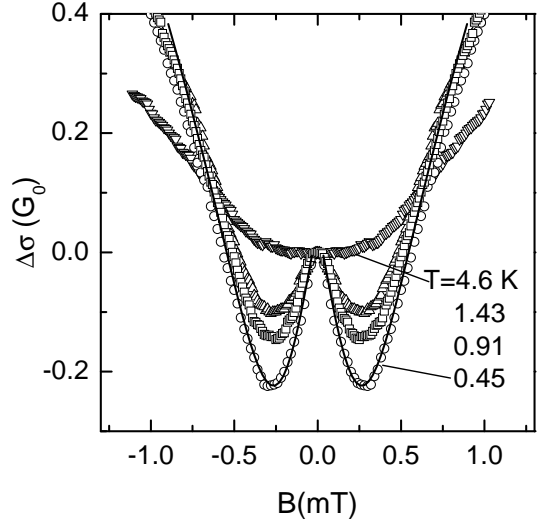}
\caption{The low-field magnetoconductance taken for different
temperatures for the structure 3635 at $V_g=0$. Symbols are the
experimental data. Solid line is the best fit by ILP-formula
\protect\cite{ILP} with the parameters $B_{tr}=2.4$~mT,
$\tau_\phi=7.2\times10^{-11}$~s, and
$\tau_s=\tau^\prime_s=0.88\times10^{-11}$~s.} \label{f2}
\end{figure}

Figure \ref{f2} shows the low-field magnetoconductivity
$\sigma(B)=\rho_{xx}^{-1}(B)$ measured for the structure 3635 as a
function of magnetic field $B$ for different temperatures.  The
antilocalization minimum in low fields, which magnitude decreases
with temperature increase, is clearly seen.

\begin{figure}[t]
\includegraphics[width=14cm,clip=true]{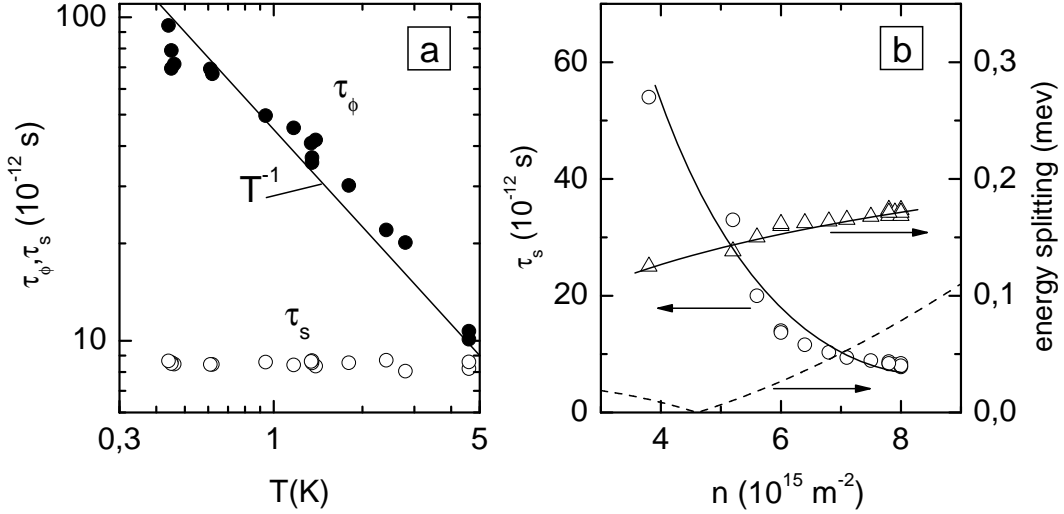}
\caption{(a) The temperature dependence of $\tau_\phi$ and
$\tau_s$ found from the fitting procedure for structure 3635 at
$V_g=0$. (b) The value of $\tau_s$ and energy splitting
$\Delta=\hbar/\sqrt{2\tau\tau_s}$ as  functions of electron
density controlled by the gate voltage for structure 3635. Symbols
are the experimental data, dashed line is result of
self-consistent calculations for the Rashba spin-orbit splitting
$\Omega_R$, solid lines are provided as a guide for the eye.
}\label{f3}
\end{figure}

The data treatment was performed using the model developed by
Iordanskii, Lyanda-Geller, and Pikus (ILP),\cite{ILP} which took
into account both linear and cubic in $k$ terms in the energy
spectrum. Within framework of this theory four characteristic
magnetic fields describe the behavior of interference induced
magnetoresistance in the presence of spin relaxation. They are two
standard parameters: $B_{tr}=\hbar/(4eD\tau)$ and
$B_{\phi}=\hbar/(4eD\tau_\phi)$, where $D$ is the diffusion
coefficient, $\tau$ and $\tau_\phi$ are the momentum and phase
relaxation time, respectively, and two additional parameters
relevant to spin relaxation: $B_{so}=\hbar/(4eD\tau_s)$ and
$B^\prime_{so}=\hbar/(4eD\tau^\prime_s)$ with
$\tau_s=\left(2\Omega^2 \tau_1+2\Omega_3^2\tau_3\right)^{-1}$, and
$\tau^\prime_s=\left(2\Omega^2 \tau_1\right)^{-1}$. Here, $\Omega$
is equal to $\Omega_R$ or $\Omega_1$, and $\tau_1$ and $\tau_3$
are the transport relaxation times introduced as
$\tau_i^{-1}=\int(1-\cos i\theta)W(\theta)d\theta$, where
$W(\theta)$ is the probability of the scattering per angle
$\theta$ per unit time (note, $\tau_1=\tau$). The value of
$B_{tr}$ was found in the ordinary way using the results of Hall
and conductivity measurements. The other three parameters are in
principle free and can be used as the fitting ones. However, our
analysis shows that the experimental data for structure 3635 are
in excellent agreement with ILP-expression under the condition
$B_{so}= B^\prime_{so}$ [see Fig.~\ref{f2}]. The temperature
dependences of $\tau_\phi$ and $\tau_s$ for $V_g=0$ are presented
in Fig.~\ref{f3}(a). It is seen that the fitting parameter
$\tau_s$ is independent of $T$ that corresponds to the
Dyakonov-Perel mechanism of spin relaxation.\cite{MinDP} The
temperature dependence of $\tau_\phi$ is close to the $T^{-1}$-law
predicted by conventional theory for the temperature dependence of
the phase relaxation time when the dephasing is caused by
inelasticity of electron-electron interaction.  Such a natural
behavior of $\tau_s$ and $\tau_\phi$ with temperature allows us to
believe that we have experimentally obtained the phase and spin
relaxation times.

The fact that the fitting procedure does not fail at $B_{so}=
B^\prime_{so}$ means that the main contribution to
antilocalization comes from the linear in $k$ term. In order to
understand what mechanism, Rashba or Dresselhaus, is responsible
for the energy splitting resulting in the spin relaxation, we have
performed self-consistent calculations of the electron energy
spectrum within framework of $kP$-model.

Let us start with consideration of the Rashba mechanism. It is not
so difficult to obtain rather reliable theoretical results in this
case because the heterostructure geometry  and all the material
parameters are known. Figure~\ref{f3}(b) shows the electron
density dependences of the energy splitting obtained
experimentally as $\Delta=\hbar/\sqrt{2\tau\tau_s}$ together with
the results of calculations of $\Omega_R$. It is seen that the
theoretical values of the splitting is strongly less than
experimental ones in whole electron density range. From the first
sight it seems unnatural because the structure 3635 was grown with
strong In-gradient within the quantum well. However our analysis
shows that the contribution of inclined part of the well (see
Fig.~\ref{f1}(a)) to the splitting is compensated by that of the
right-side abrupt InGaAs/GaAs interface. This effect takes place
due to insufficient height of the right-side barrier that results
in strong penetration of the wave function into it. Thus, we
conclude that not Rashba effect is responsible for the
antilocalization in the structure 3635.

\begin{figure}[t]
\includegraphics[width=14cm,clip=true]{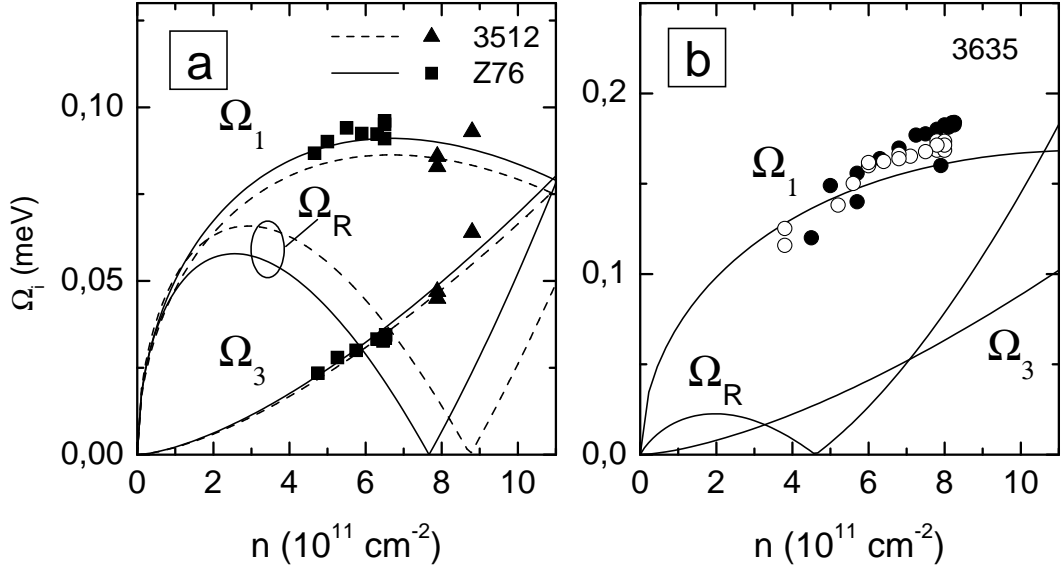}
\caption{The values of $\Omega_R$, $\Omega_1$, and $\Omega_3$ as
functions of electron density for structures 3512, Z76 (a), and
3635 (b). Circles are the experimental data for $T=0.45$ K (open
symbols) and $1.5$~K (solid symbols), lines are the result of
self-consistent calculations.} \label{f4}
\end{figure}

To calculate the energy splitting caused by the Dresselhaus effect
one needs to know the value of parameter $\gamma$. This parameter
is expressed through seven (!) band parameters of the volume
material.\cite{Knap} Not all of them are known with high enough
accuracy, therefore, we have obtained the parameter $\gamma$
experimentally. For this purpose we have used the experimental
results obtained for the structures Z76 and 3512. For these
structures it was impossible to fit satisfactorily the
experimental $\sigma$-versus-$B$ curves by ILP-formula under the
condition $B_{so}=B^\prime_{so}$. This means that both linear and
cubic in $k$ terms contribute to antilocalization in these
structures. To simplify the fitting procedure in this case we did
not consider the parameters $B_{so}$ and $B^\prime_{so}$ as
independent ones. We supposed that they were interrelated via the
energy spectrum and scattering anisotropy in accordance with their
definition: $B_{so}/
B^\prime_{so}=1+\Omega_3^2\tau_3/(\Omega^2\tau_1)$, where $\Omega$
stands for $\Omega_1$ or $\Omega_R$. The ratio $\Omega_3/\Omega$
was chosen equal to that found from the calculation of the energy
spectrum. The ratio $\tau_3/\tau_1$ was from the range $0.5...0.7$
depending on the scattering anisotropy which in its turn depended
on the electron density. The best fit of $\sigma$-versus-$B$
curves has been achieved when $\Omega=\Omega_1$ that points to the
fact that the main contribution to the spin relaxation comes from
the Dresselhaus effect. The final results of such a data
processing for structures Z76 and 3512 are shown in
Fig.~\ref{f4}(a) by symbols. Comparing the experimental and
calculated electron density dependences of $\Omega_1$ and
$\Omega_3$ we have found that the best agreement between theory
and experiment is achieved for $\gamma=18$~eV\,\AA$^3$ (lines in
Fig.~\ref{f4}(a)). An overall error in determination of $\gamma$
can be estimated as $\pm 2$~eV\,\AA$^3$. Finally we can return to
the results obtained for the structure 3635. Figure~\ref{f4}(b)
demonstrates satisfactory agreement between experimental data and
results calculated with just the same value of
$\gamma=18$~eV\,\AA$^3$.

Comparing Figs.~\ref{f4}(a) and \ref{f4}(b) one can see that the
$\Omega_3$ to $\Omega_1$ ratio is close for both symmetric and
asymmetric structures. The reasonable question arises in this
connection: why does not the cubic in $k$ splitting contribute to
the antilocalization in structure 3635? From our point of view one
of possible reason is that the doped layer is more spaced from the
quantum well in this structure as compared to the other
structures. This results in smoother scattering potential and
stronger scattering anisotropy which in its turn determines the
relative contribution of cubic and linear terms in
antilocalization via $\tau_3$ to $\tau_1$ ratio.

\section{Conclusion}
We have investigated the antilocalization caused by the spin-orbit
interaction in GaAs/In$_x$Ga$_{1-x}$As/GaAs quantum wells with and
without gradient of indium content in growth direction. It has
been found that the Dresselhaus effect is responsible for
antilocalization in both types of the heterostructures. The
comparison of experimental data with the results of
self-consistent calculations allows us to determine reliably the
value of $\gamma$ parameter: $\gamma=(18\pm 2)$~eV\AA$^3$.


This work was supported in part by the RFBR through Grants No.
01-02-17003, No. 03-02-16150, and No.~03-02-06025, the INTAS
through Grant No. 1B290, the CRDF through Grant No. REC-005, the
Program {\it University of Russia} through Grant No.
UR.06.01.002. \vspace{2mm} \\

\end{document}